\documentclass[aps,prd,reprint,groupedaddress,nofootinbib]{revtex4-1}

\usepackage{graphicx}
\usepackage{multirow} 
\usepackage{amssymb}
\usepackage{amsmath}
\usepackage{enumerate}
\usepackage{todonotes}
\usepackage{hyperref}

\begin{document}


\title{Erratum: A Comprehensive Approach to Tau-Lepton Production by High-Energy Tau Neutrinos 
Propagating Through Earth\\Phys. Rev. D 97, 023021 (2018) } 
\author{Jaime Alvarez-Mu\~niz$^1$, Washington R. Carvalho Jr.$^{2,1}$, Austin L. Cummings$^{3}$, K\'evin Payet$^4$, Andr\'es Romero-Wolf$^5$, Harm Schoorlemmer$^6$, Enrique Zas$^1$}
\affiliation{$^1$Departamento de F\'\i sica de Part\'\i culas \& Instituto Galego de F\'\i sica de Altas Enerx\'\i as, Univ. de Santiago de Compostela, 15782 Santiago de Compostela, Spain\\
$^2$Departamento de F\'\i sica, Universidade de S\~ao Paulo, S\~ao Paulo, Brazil\\
$^3$ Gran Sasso Science Institute, School of Advanced Studies, L'Aquila, Italy\\
$^4$ Universit\'e Joseph Fourier (Grenoble I); Currently at La Javaness, 75010, Paris, France\\
$^5$ Jet Propulsion Laboratory, California Institute of Technology, Pasadena, CA 91109, USA\\
$^6$ Max-Planck-Institut f\"ur Kernphysik, 69117, Heidelberg, Germany}

\maketitle



We report an error found during independent review of the publicly available code\footnote{\url{https://github.com/harmscho/NuTauSim}} that forms the basis of this publication. The error in the code was in tracking the density of the medium during particle propagation. After the first interaction, the code was referencing the depth of penetration back to the surface of the Earth rather than the location of the last interaction. The results were obtained using densities that were systematically underestimated when the particle was traversing the inner layers of the Earth by assigning the density of either ice or bedrock, depending on the particle energy or ice thickness of the simulation. This error was fixed and the repository updated on September 29, 2018.

The main impact of the coding error is that the $\tau$ lepton exit probability $P_\mathrm{exit}$ was being overestimated for emergence angles greater than the value corresponding to the interface between the outermost layer of the Earth model and the layer below. In the case of a layer of ice with 4~km thickness, the impact is on emergence angles $\bar\theta>2^{\circ}$. All of the figures that used simulation results in the original paper are reproduced here with the same Figure number to facilitate comparison. In the updated Figure~\ref{fig:tau_exit_prob}, $P_\mathrm{exit}$ is unaffected for $\bar\theta\leq2^{\circ}$ but suppressed compared to the original result for $\bar\theta>2^{\circ}$. The suppression increases with emergence angle and depends on the tau neutrino energy $E_{\nu_\tau}$, reaching up to a factor of $\sim$ 5 at $\bar\theta=30^\circ$. 
%
%
The shape of the distribution of exiting $\tau$-lepton energies (gray bands in Figure~\ref{fig:tau_energy_distrib}) did not significantly change after the update. The reason for this is that exiting $\tau$ leptons are produced near the surface and mostly propagate within the ice because their their range is limited to $\sim$50 km at $10^{21}$~eV and $\sim$5 km at $10^{17}$~eV. The error in the code was mis-assigning the density only at layers below the first surface layer. While this could modify the shape of the exiting $\tau$ lepton energy distribution for high emergence angles, it is not very relevant because $P_\mathrm{exit}$ is already highly suppressed.
The mean number of CC and NC interactions and tau decays, shown in Figure~\ref{fig:tau_interactions}, is unaffected for $\bar\theta\leq2^\circ$ but shows a slight increase for $\bar\theta>2^{\circ}$, as expected from an increase in density after the coding error fix. 

The general conclusions about the effect of $\nu_\tau$ regeneration are not modified. In the updated Figure~\ref{fig:tau_energy_distrib_regen}, the effect of not including the effects of regeneration is still to significantly suppress $P_\mathrm{exit}$ for $\bar\theta>2^{\circ}$, increasingly so with higher emergence angle. The updated figure includes additional suppression due to the coding error discussed above. The distributions of $\tau$ lepton energies shown in Figure~\ref{fig:tau_energy_distrib_regen} are not significantly modified after the coding error fix for the same reasons discussed the previous paragraph. 

The biggest impact on the code fix is on the behavior of $P_\mathrm{exit}$ with ice thickness, as shown in Figure~\ref{fig:tau_exit_prob_depth}. This effect was originally studied in~\cite{Palomares-Ruiz_2006}. Prior to the code fix, we had concluded that a layer of ice increased $P_\mathrm{exit}$ at emergence angles 
$\bar\theta\gtrsim0.3^{\circ}-1.0^{\circ}$ (depending on energy) for neutrino energies $E_{\nu_\tau}>3\times 10^{18}$~eV. With the updated simulation, this range of emergence angles where $P_\mathrm{exit}$ benefits from a layer of ice is restricted to $0.3^\circ\lesssim\bar{\theta}\lesssim 3^{\circ}$. Note that this range of emergence angles corresponds to the majority of the area observed by a detector at altitude~\cite{Romero-Wolf_2018}. The reason for this change is that the densities at higher emergence angles were systematically being underestimated resulting in suppression of the neutrino interaction probability in the subsurface rock layers. At energies $E_{\nu_\tau}<3\times 10^{18}$~eV, the code fix results in $P_\mathrm{exit}$ being higher for bare rock than ice at all emergence angles. For $E_{\nu_\tau}=10^{17}$~eV, we find that bare rock has approximately twice the $P_\mathrm{exit}$ than for a layer of ice. This behavior is explained by the probability of neutrino interaction being higher in rock by a factor of $\sim2.8$ while the tau range being only $\sim30\%$ longer in ice. 

To better understand the behavior at lower energies, we ran a set of simulations with $E_{\nu_\tau}=10^{17}$~eV and constant $\bar\theta=10^\circ$ for various ice thicknesses (Figure~\ref{fig:Ice_Thickness_Fine}). 
%
%
For ice thickness $D\geq1$~km, $P_\mathrm{exit}$ is constant. This is because leptons produced in rock are mostly absorbed in the ice and only leptons produced in the ice $\sim$3~km away from the surface (the $\tau$-lepton range in this energy scale) are able to exit. Note that for this geometry ($\bar\theta=10^\circ$), the track length $L$ in ice, after traversing rock, is $L\geq6$~km for $D\geq1$~km, and increases with increasing ice thickness.
In this case, $\tau$ leptons produced by neutrino interactions in the ice dominate the contribution to $P_\mathrm{exit}$. For ice thickness $\lesssim1$~km, the range of $\tau$ leptons produced by interactions in rock (where the $\nu_\tau$ interaction probability is higher) is sufficiently large that they have a high probability of escaping the ice layer into the atmosphere and therefore contribute significantly to $P_\mathrm{exit}$ compared to neutrino interactions in the ice layer. As the ice thickness is reduced to $\sim0.01$~km, the fraction of neutrinos interacting in ice compared to rock is negligible thereby making $P_\mathrm{exit}$ approximately constant for ice thickness $\lesssim0.01$~km. 

The relative differences between cross-section and $\tau$-lepton energy loss models (Figure~\ref{fig:tau_exit_prob_models}) does not result in any significant changes after the code fix other than the behavior of $P_\mathrm{exit}$ vs emergence angle already discussed in Figure~\ref{fig:tau_exit_prob}. The distribution of exiting $\tau$ leptons for the various models, shown in Figure~\ref{fig:tau_energy_models}, also does not show significant differences after the code fix. 

The exiting $\tau$ lepton flux resulting from an incoming $\nu_\tau$, shown in Figure~\ref{fig:Kotera_tau_flux}, behaves as expected based on the discussion above: for emergence angle $\bar\theta=1^\circ$ the results are unchanged while for $5^\circ$ and $10^\circ$, the spectra retain the same shape but with a lower integrated flux by a factor of $\sim 3$, consistent with the change in $P_\mathrm{exit}$ in the old and new Figure~\ref{fig:tau_exit_prob}. The fluxes with interaction histories shown in Figure~\ref{fig:Channels_RockIce} are unchanged after the code fix for $\bar\theta=0.2^\circ$,  $1.0^\circ$, and $2.0^\circ$, as expected. For $\bar\theta=5.0^\circ$, the exiting $\tau$ lepton fluxes for bare rock (dashed lines) are not modified since the subsurface layers do not significantly change in density for these trajectories. For the 4~km thick ice layer, however, the flux curves are suppressed by a factor of 2, which is consistent with the change in $P_\mathrm{exit}$ in Figure~\ref{fig:tau_exit_prob} after the code fix. The contribution from trajectories that underwent exactly one CC interaction (green), corresponding to events with no $\nu_\tau$ regeneration, was not changed since these tend to occur near the surface. The contribution from trajectories that had at least one NC interaction (red) or at least one $\tau$ lepton decay (black) are suppressed by a factor of $\sim3$, which is consistent with the coding error that was incorrectly assigning the surface ice density rather than the subsurface bedrock density.

Making the code publicly available has succeeded at motivating independent review and improving the quality of the simulations. 
The code fix has left most conclusions of the original paper qualitatively unchanged with one slight modification: the benefit of having a layer of ice for $E_{\nu_\tau}>3\times10^{18}$~eV is limited to emergence angles from $\sim0.3^\circ$ to $\sim3.0^\circ$ and not to higher emergence angles as originally stated. 


\setcounter{figure}{4}
\begin{figure}[ht]
  \centering
   \includegraphics[width=1.0\linewidth]{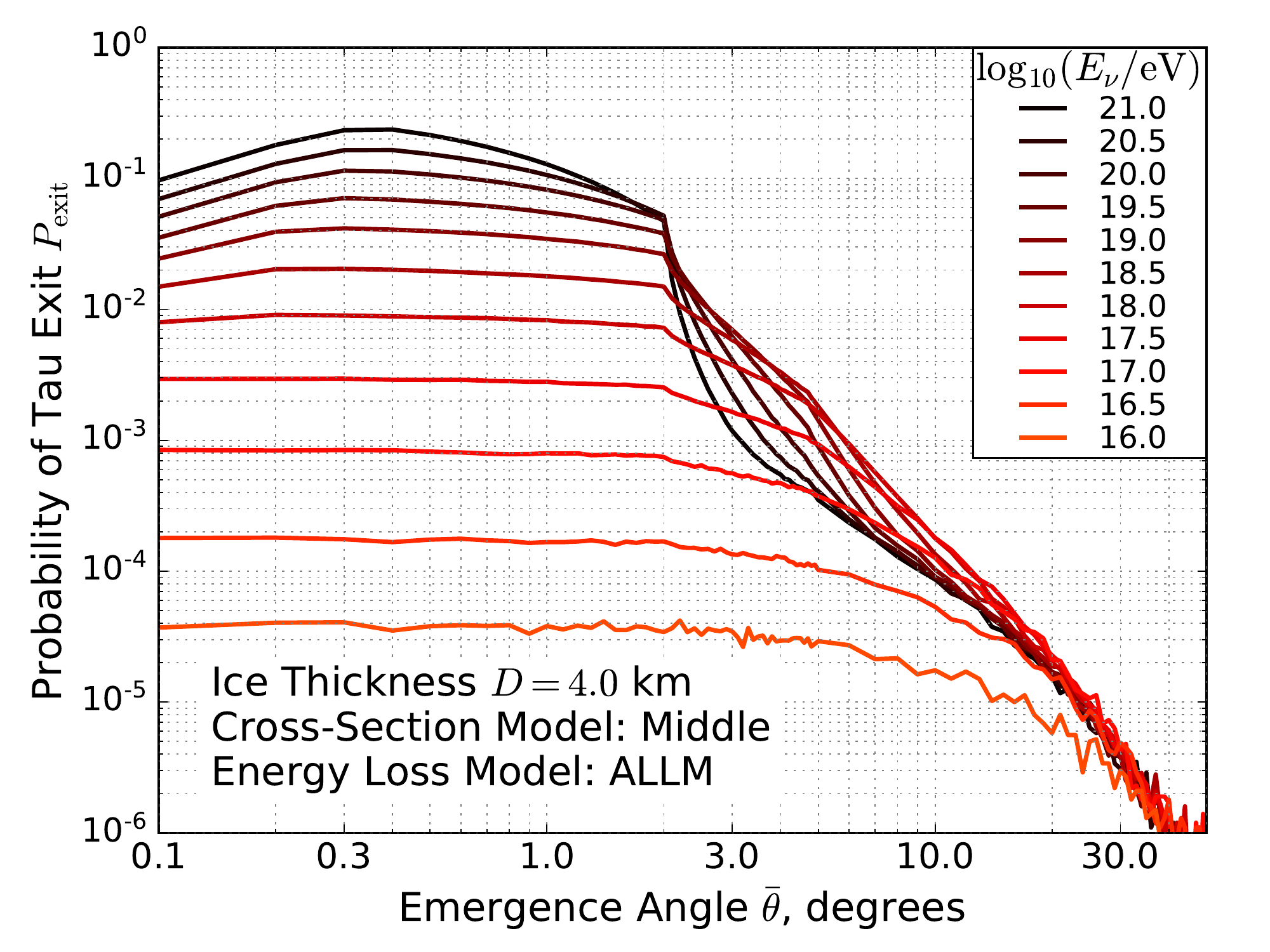} %
   \caption{The probability $P_\mathrm{exit}$ that a $\tau$ lepton exits the Earth's surface for emergence angles between 0.1$^{\circ}$ (Earth skimming) and 50$^{\circ}$ given a 4 km~thick layer of ice with standard cross-sections and energy-loss models. The feature at emergence angle of $2^{\circ}$ corresponds to the trajectory tangential to the rock layer beneath the 4~km thick layer of ice. }
   \label{fig:tau_exit_prob}
\end{figure}

\begin{figure}[!t]
  \centering
   \includegraphics[width=1.0\linewidth]{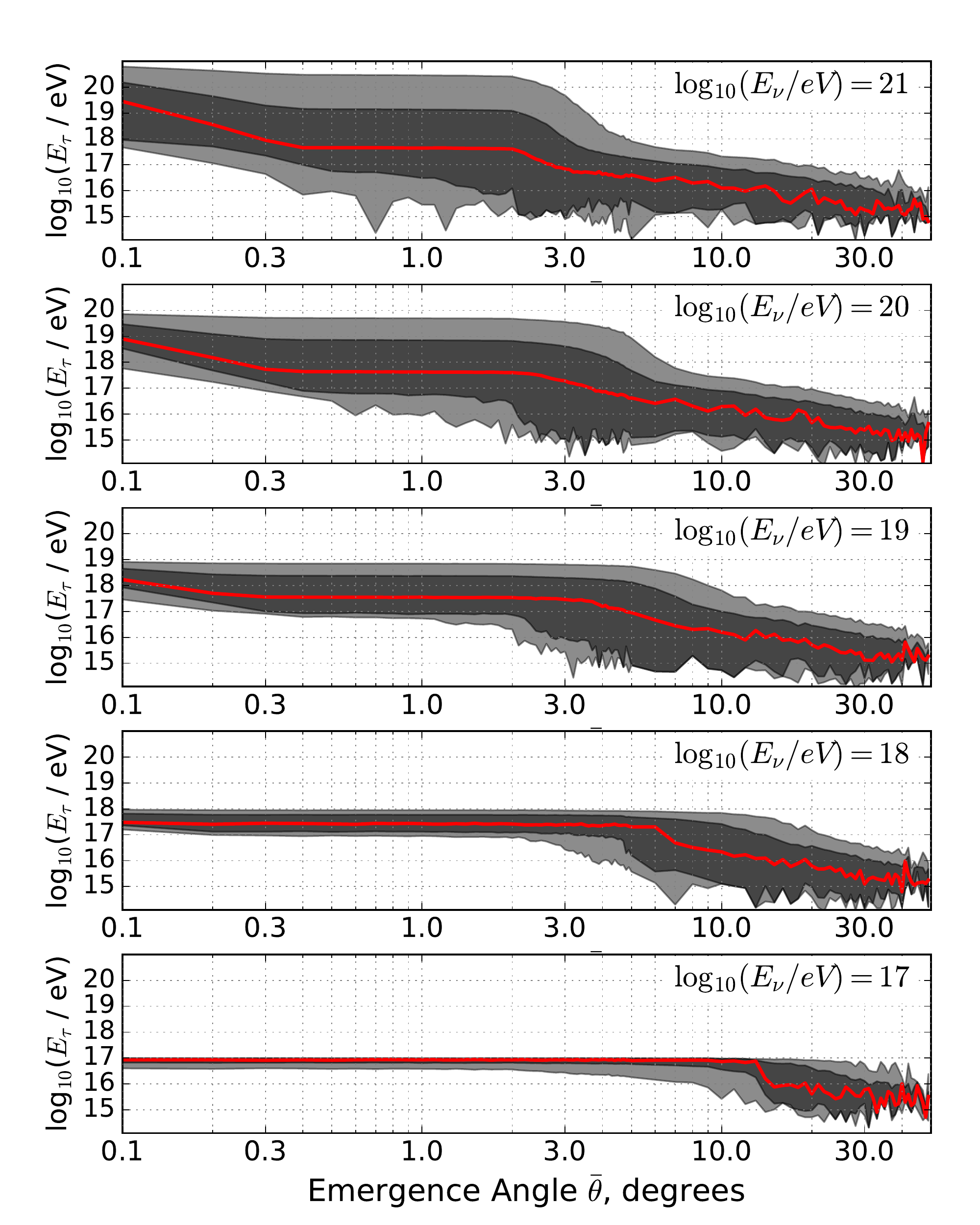} %
   \caption{The exiting $\tau$ lepton energies corresponding to some of the energies shown in Figure~\ref{fig:tau_exit_prob}. The red line shows the most probable exiting tau lepton energy. The dark (light) gray band shows the 68\% (95\%) densest probability interval. The features in the curves are caused by regions where various interaction processes dominate. See Figure~\ref{fig:tau_interactions} and text for details. }
   \label{fig:tau_energy_distrib}
\end{figure}

\begin{figure}[!t]
  \centering
   \includegraphics[width=1.0\linewidth]{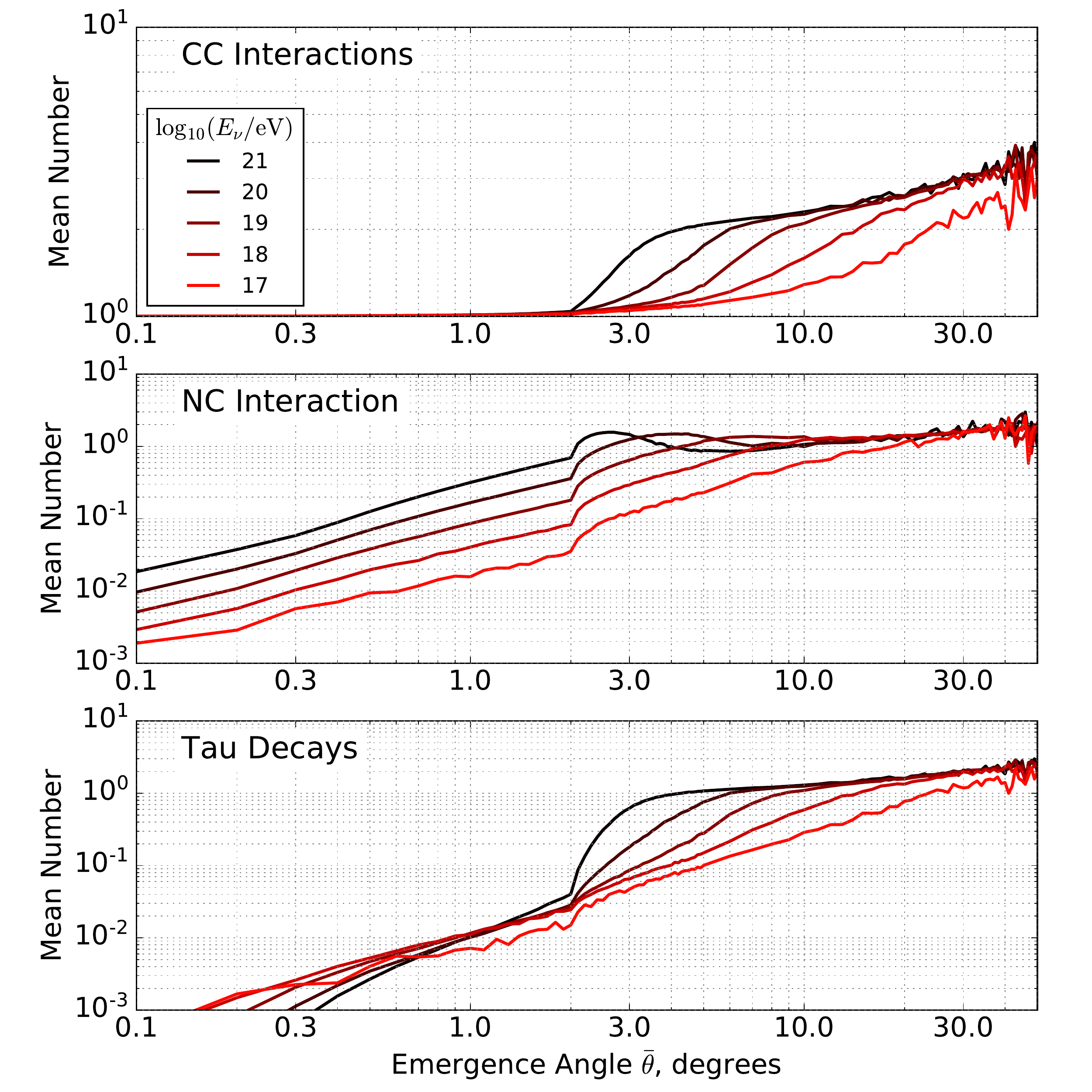} %
   \caption{The mean number of CC, NC interactions, and tau lepton decays as a
     function of emergence angle for various incident neutrino energies
     corresponding to
     Figures~\ref{fig:tau_exit_prob}~\&~\ref{fig:tau_energy_distrib}. Top
     panel: the mean number of CC interaction must be at least one since we
     are selecting for particles resulting in a $\tau$ lepton exiting the
     Earth's surface. Middle Panel: The mean number of neutral current
     interactions. The sharp transition at emergence angle $\bar \theta=2^{\circ}$ corresponds
     to the direction tangential to the subsurface rock beneath 4~km thick
     layer of ice. Bottom panel: the mean number of $\tau$ lepton decays also
     show a feature at $\bar\theta=2^{\circ}$. Note that for $\bar\theta<2^{\circ}$ the particle traverse ice only while for $\bar\theta>2^{\circ}$ the particle traverses a combination of rock and ice, which affects the behavior of $\tau$ lepton and $\nu_\tau$ interactions.}
   \label{fig:tau_interactions}
\end{figure}

\begin{figure}[!t]
  \centering
   \includegraphics[width=1.0\linewidth]{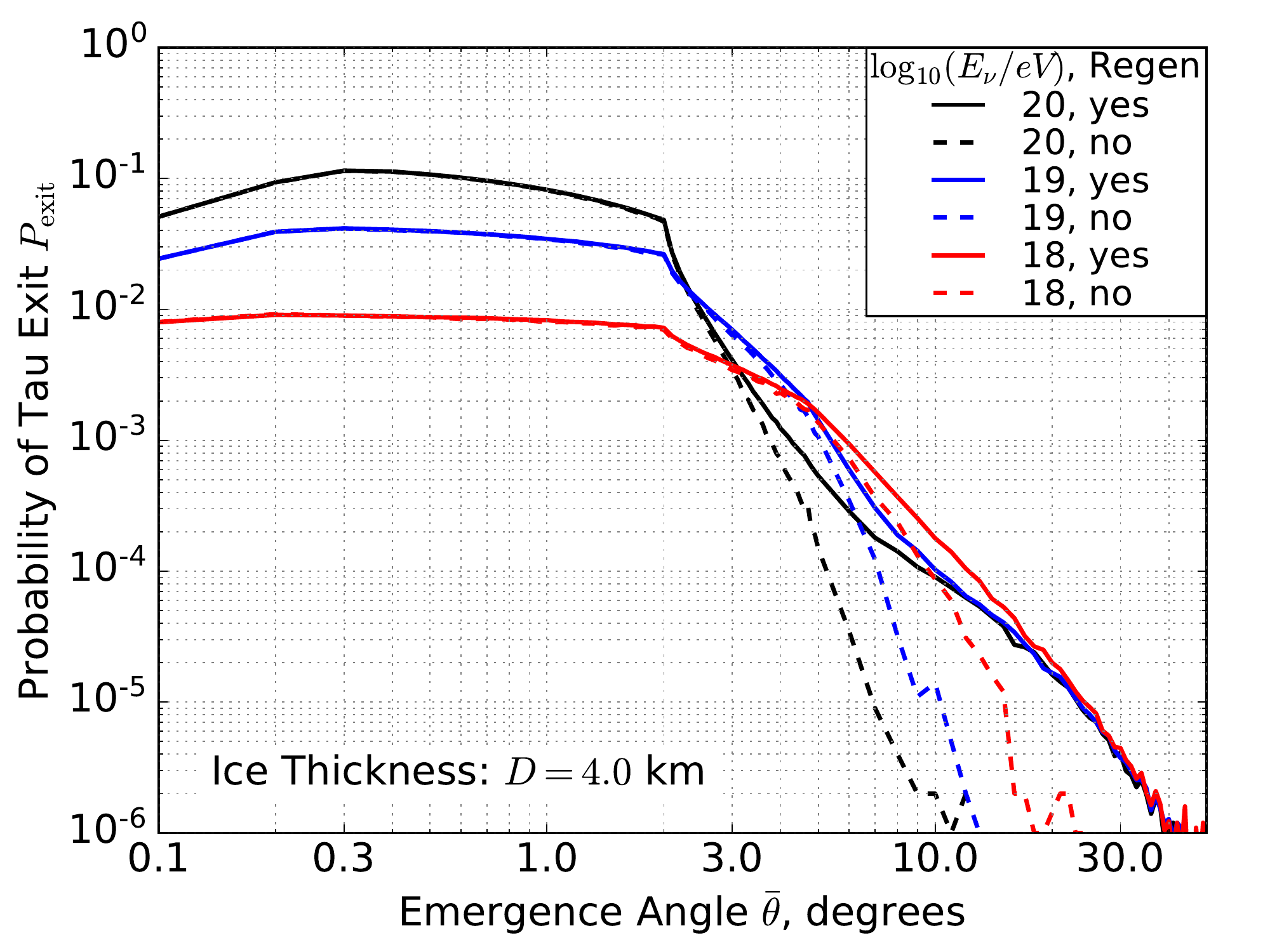} %
   \caption{The probability that a $\tau$ lepton exits Earth's surface including and excluding the effect of $\nu_\tau$ regeneration given a 4~km thick layer of ice and standard neutrino cross-section and tau lepton energy-loss models. Excluding regeneration significantly underestimates the probability of exiting $\tau$ leptons for $\bar\theta>2^{\circ}$, where the trajectories propagate through rock rather than pure ice. }
   \label{fig:tau_exit_regen}
\end{figure}

\begin{figure}[!t]
  \centering
   \includegraphics[width=1.0\linewidth]{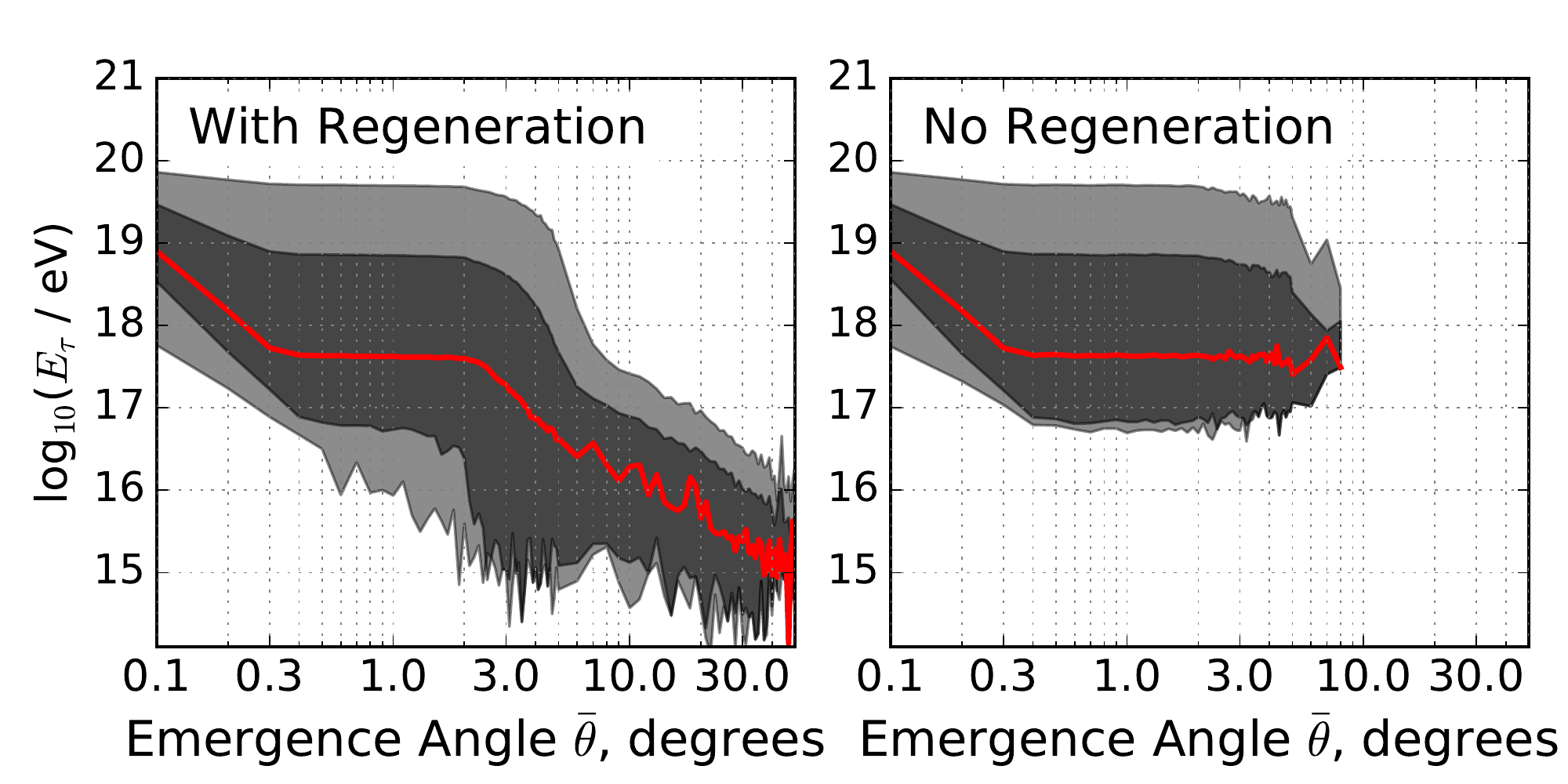} %
   \caption{The exiting $\tau$ lepton energies corresponding to $E_{\nu}=10^{20}$~eV in Figure~\ref{fig:tau_exit_regen} with and without regeneration. Excluding regeneration suppresses exiting $\tau$ leptons with energy $E_{\tau}<10^{17}$~eV.} 
   \label{fig:tau_energy_distrib_regen}
\end{figure}

\begin{figure}[t!]
  \centering
   \includegraphics[trim = {0.2cm 1.0cm 1.0cm 0.2cm}, clip, width=0.95\linewidth]{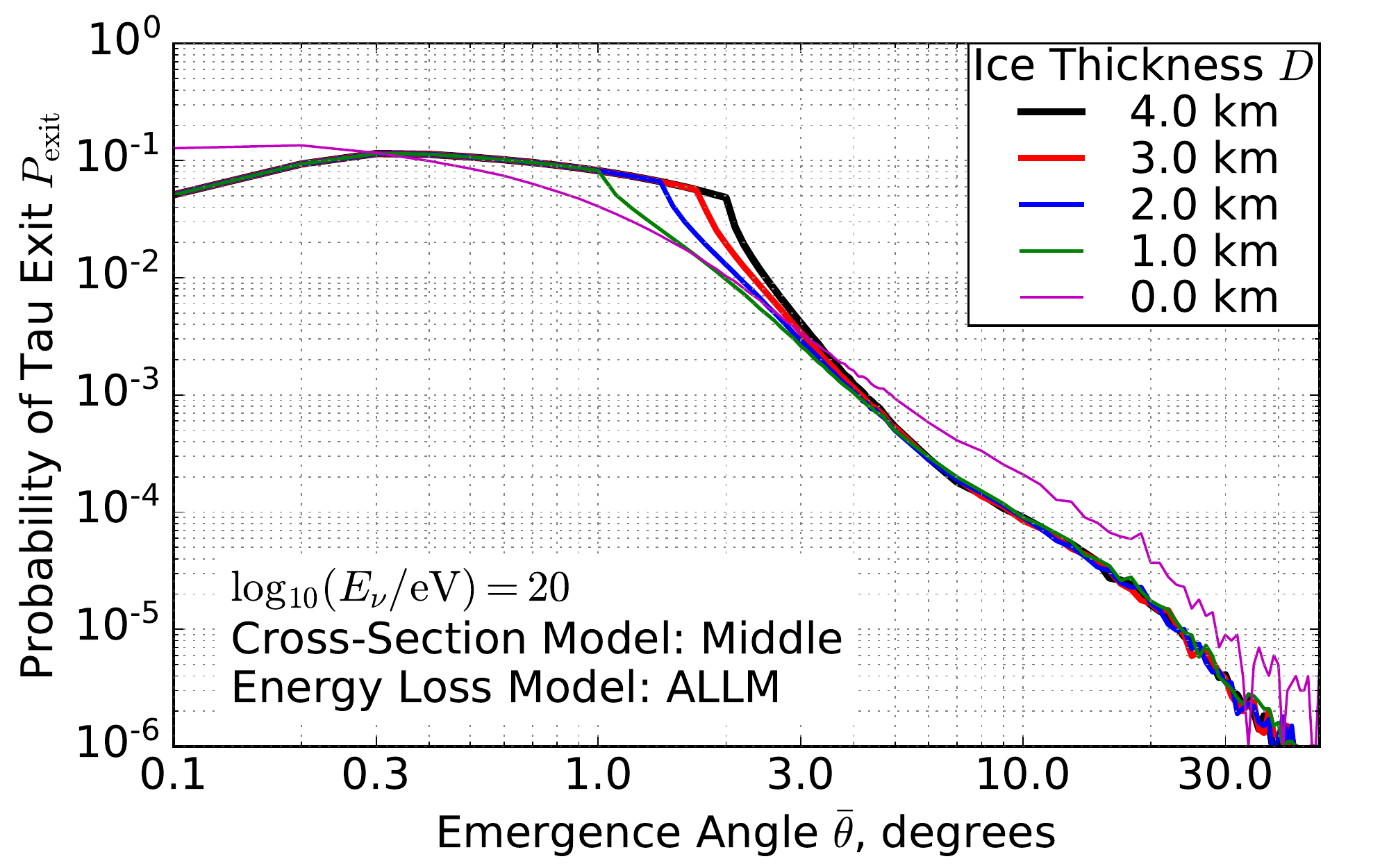} 
   \includegraphics[trim = {0.2cm 1.0cm 1.0cm 0.2cm}, clip, width=0.95\linewidth]{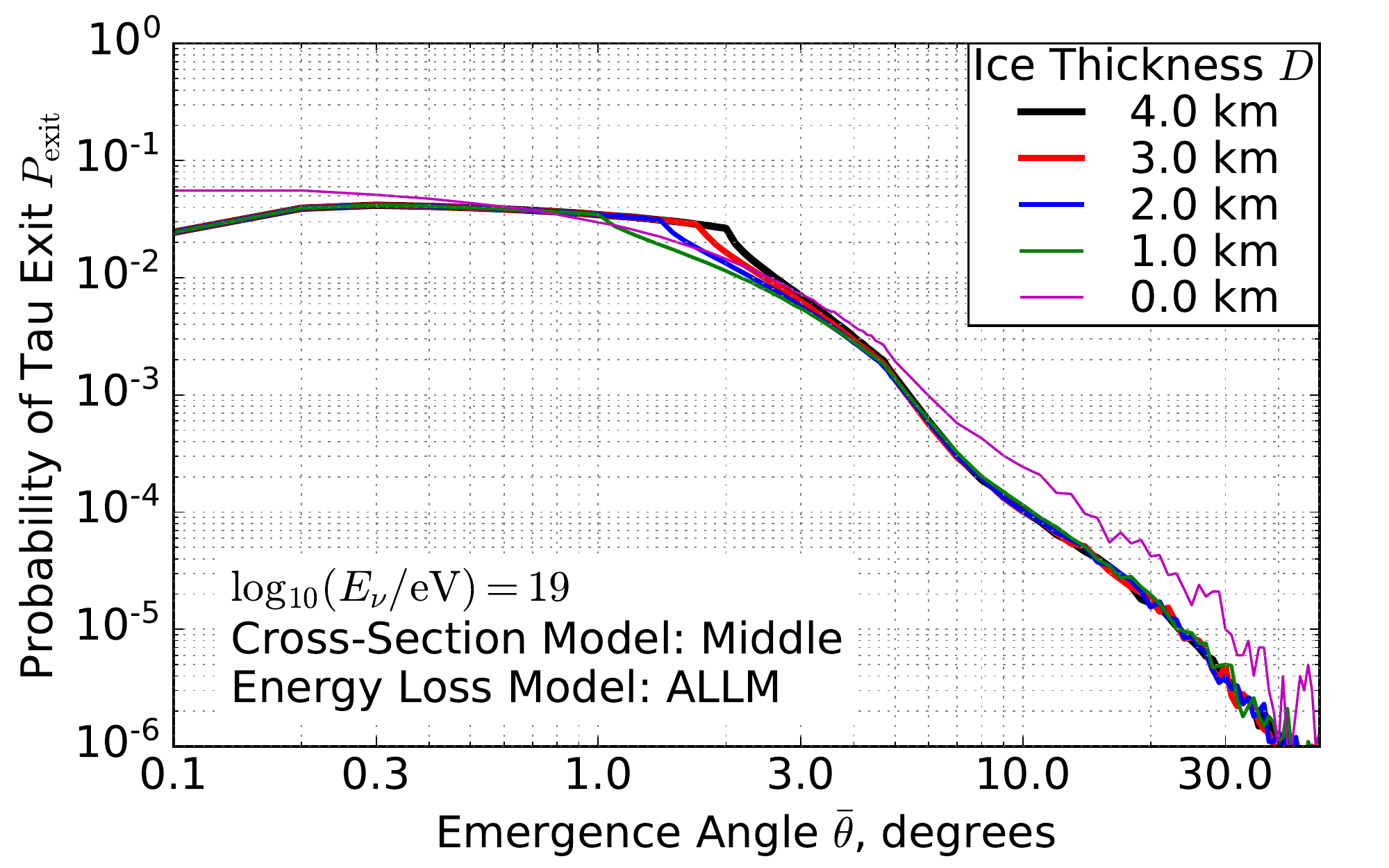} 
   \includegraphics[trim = {0.2cm 1.0cm 1.0cm 0.2cm}, clip, width=0.95\linewidth]{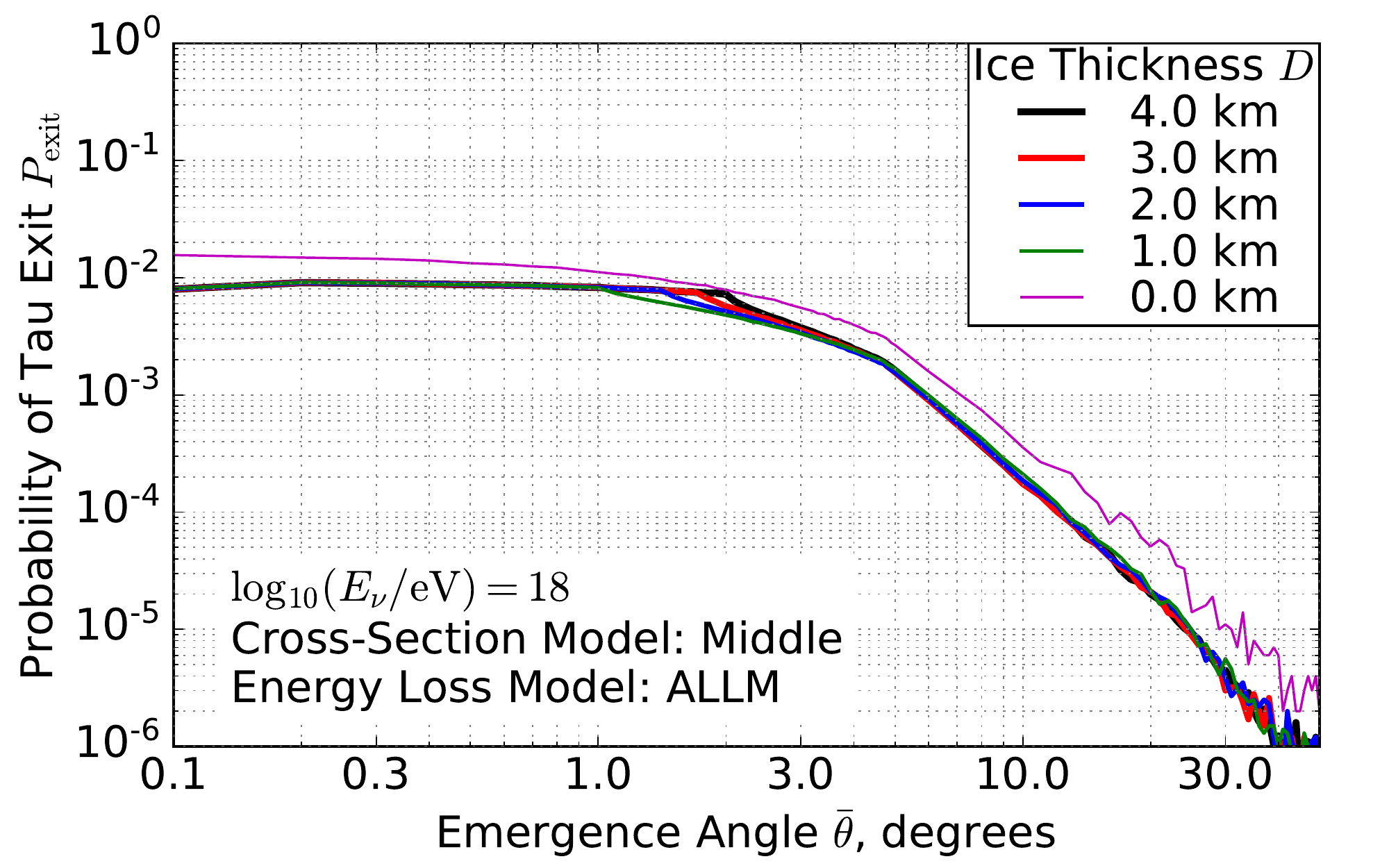} 
   \includegraphics[trim = {0.2cm 0.1cm 1.0cm 0.2cm}, clip, width=0.95\linewidth]{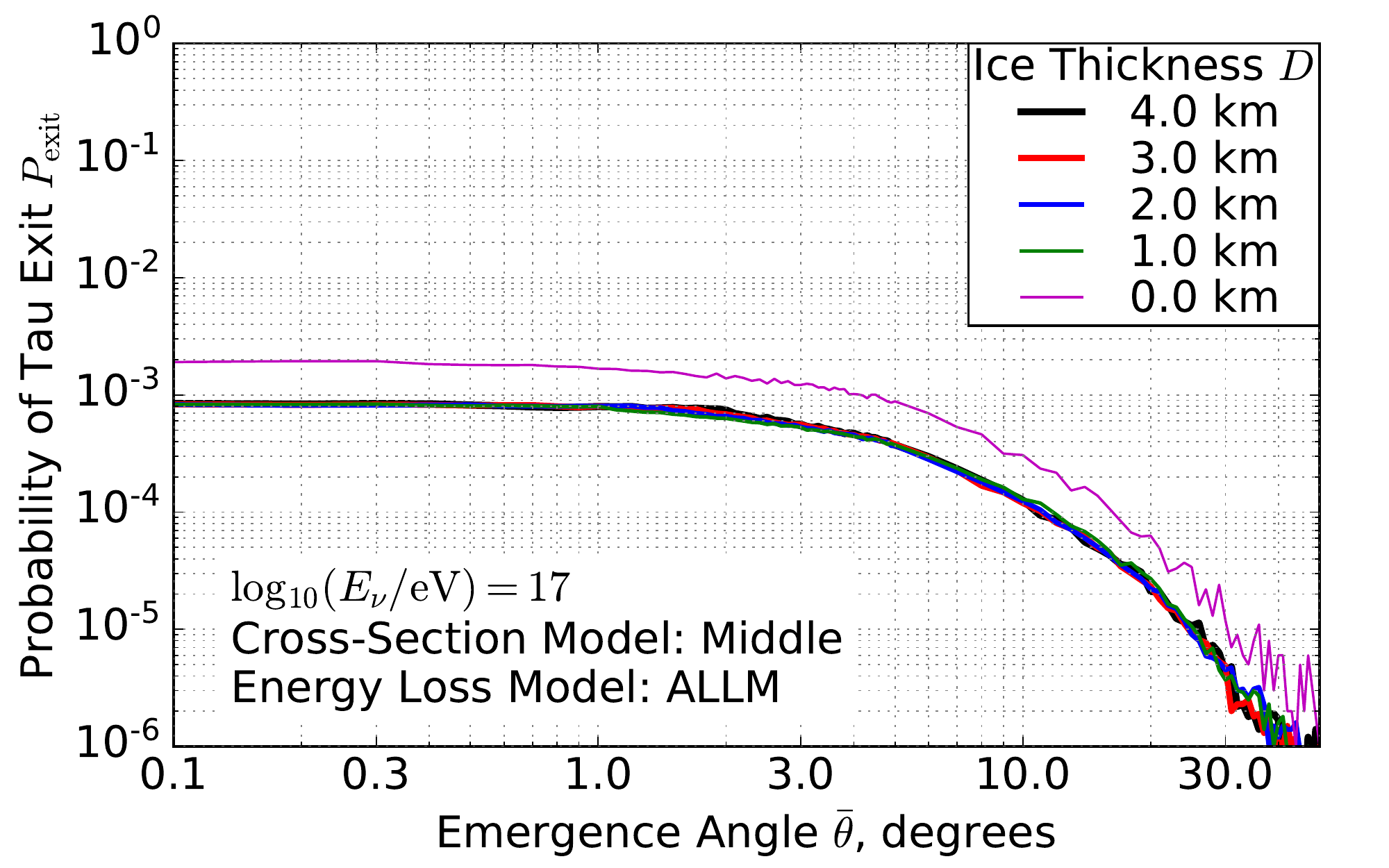} 
   \caption{The probability that a $\tau$ lepton exits the Earth's surface for various energies and ice thicknesses, including bare rock, assuming standard cross-section and energy-loss models. From top to bottom, the input neutrino energies are 10$^{20}$, 10$^{19}$, 10$^{18}$, and 10$^{17}$~eV. A layer of ice is favorable to exiting $\tau$ leptons for neutrino energies $>10^{18}$~eV while bare rock is favorable for neutrino energies $<10^{18}$~eV. See text for details.
\vspace{-30pt}
}
   \label{fig:tau_exit_prob_depth}
\end{figure}

\begin{figure}[t]
  \centering
   \includegraphics[width=1.0\linewidth]{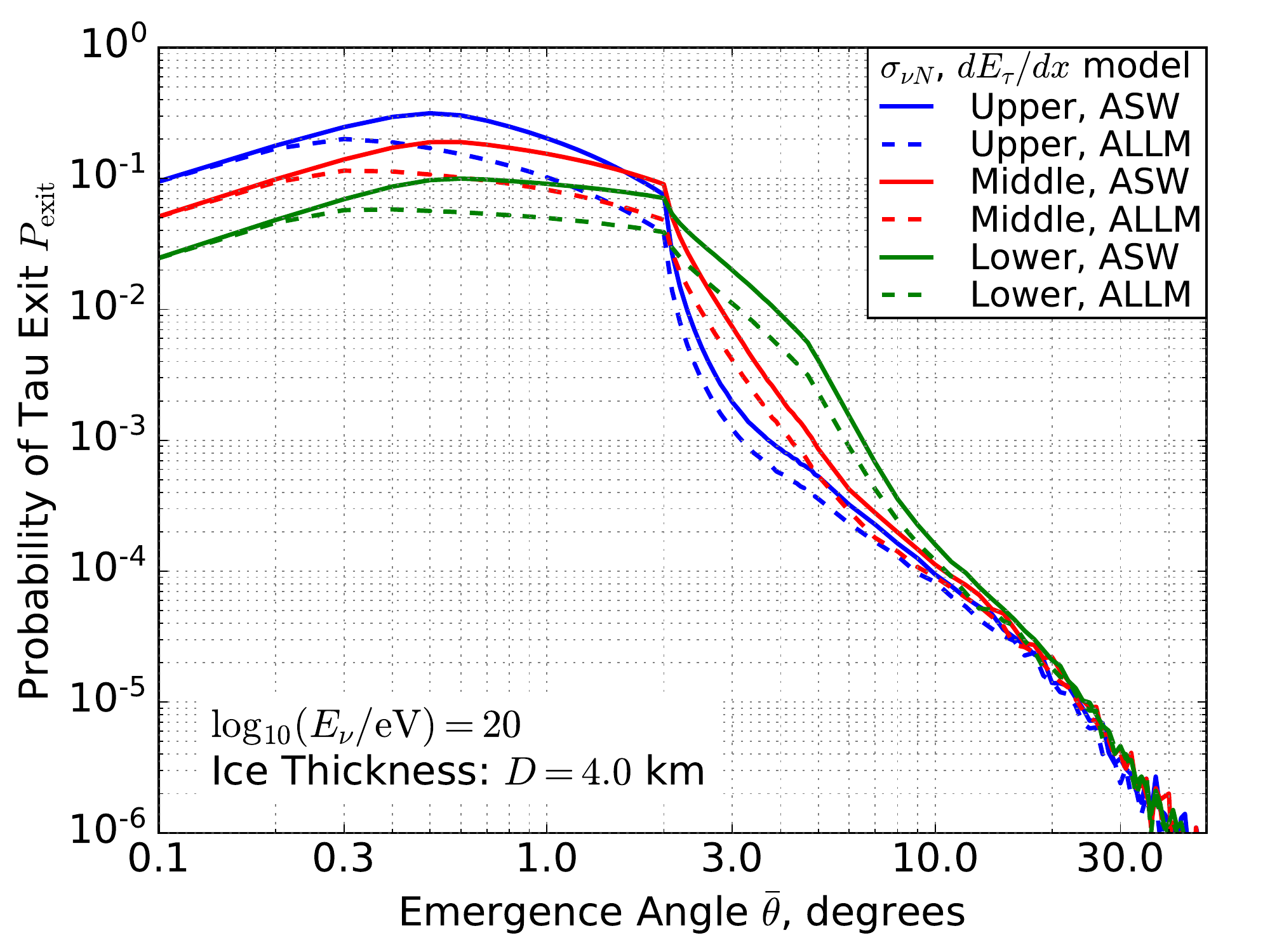} %
   \caption{The probability that a $\tau$ lepton exits the Earth's surface for various combinations of neutrino cross-section and $\tau$ lepton energy-loss models given a 4~km thick ice layer for a $10^{20}$~eV injected $\nu_\tau$. Lowering the cross-section has the general effect of reducing the $\tau$ lepton exit probability for emergence angles below where the trajectory is tangential to the subsurface rock layer while increasing the probability for larger emergence angles. The ASW energy loss rate model, which is suppressed compared to the more standard ALLM model, results in an overall increase $\tau$ lepton exit probability.}
   \label{fig:tau_exit_prob_models}
\end{figure}

\begin{figure}[t]
  \centering
   \includegraphics[width=1.0\linewidth]{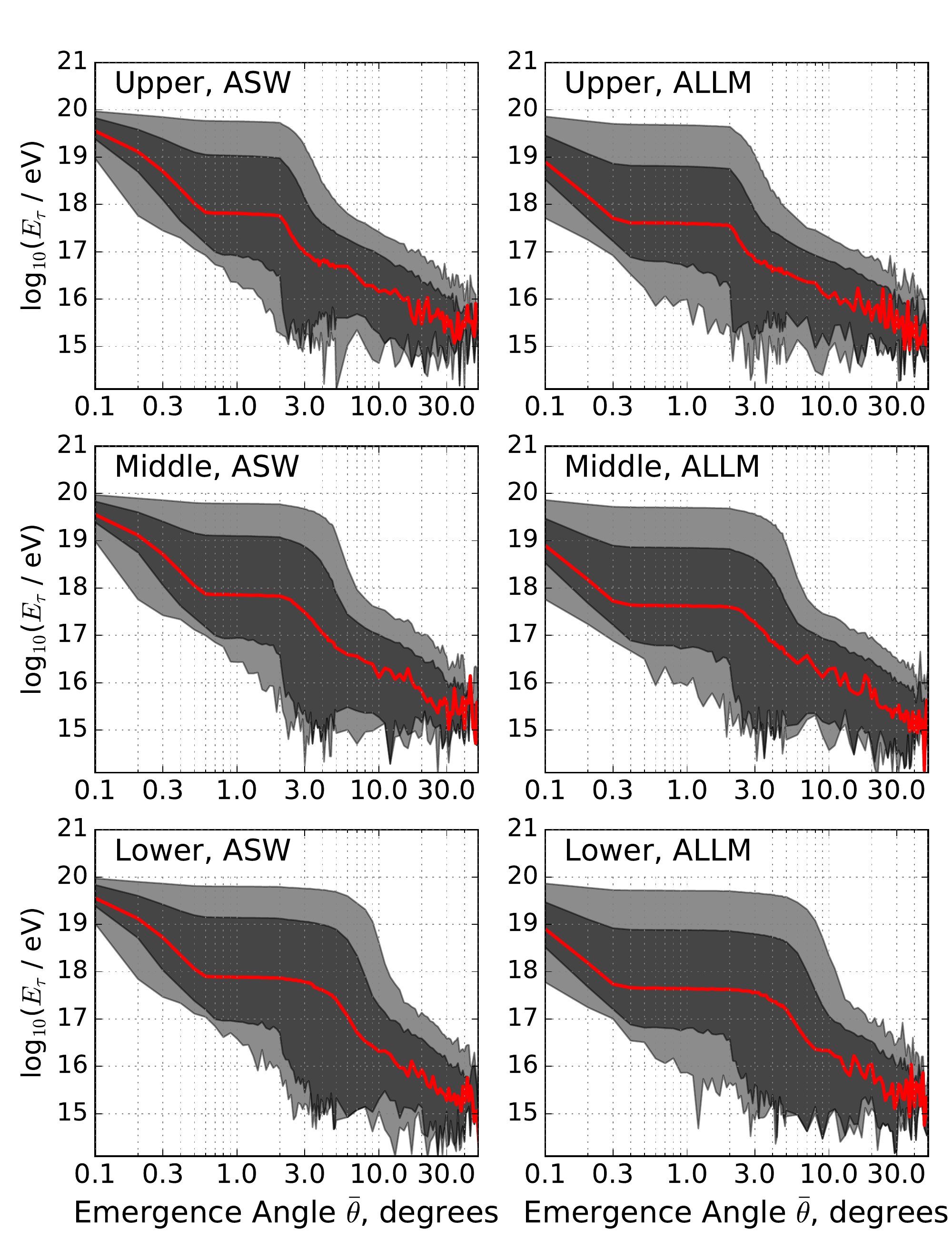} %
   \caption{The exiting $\tau$ lepton energies for various models corresponding to Figure~\ref{fig:tau_exit_prob_models}. On each panel, the cross-section model and energy loss rate models are labeled on the top left corner. The variance in exiting $\tau$ lepton energies tends to increase as the cross-section increases for trajectories that traverse mostly rock. The energy loss model changes the range of emergence angles where the most probable energies plateaus. }
   \label{fig:tau_energy_models}
\end{figure}

\begin{figure}[!t]
  \centering
   \includegraphics[width=1.0\linewidth]{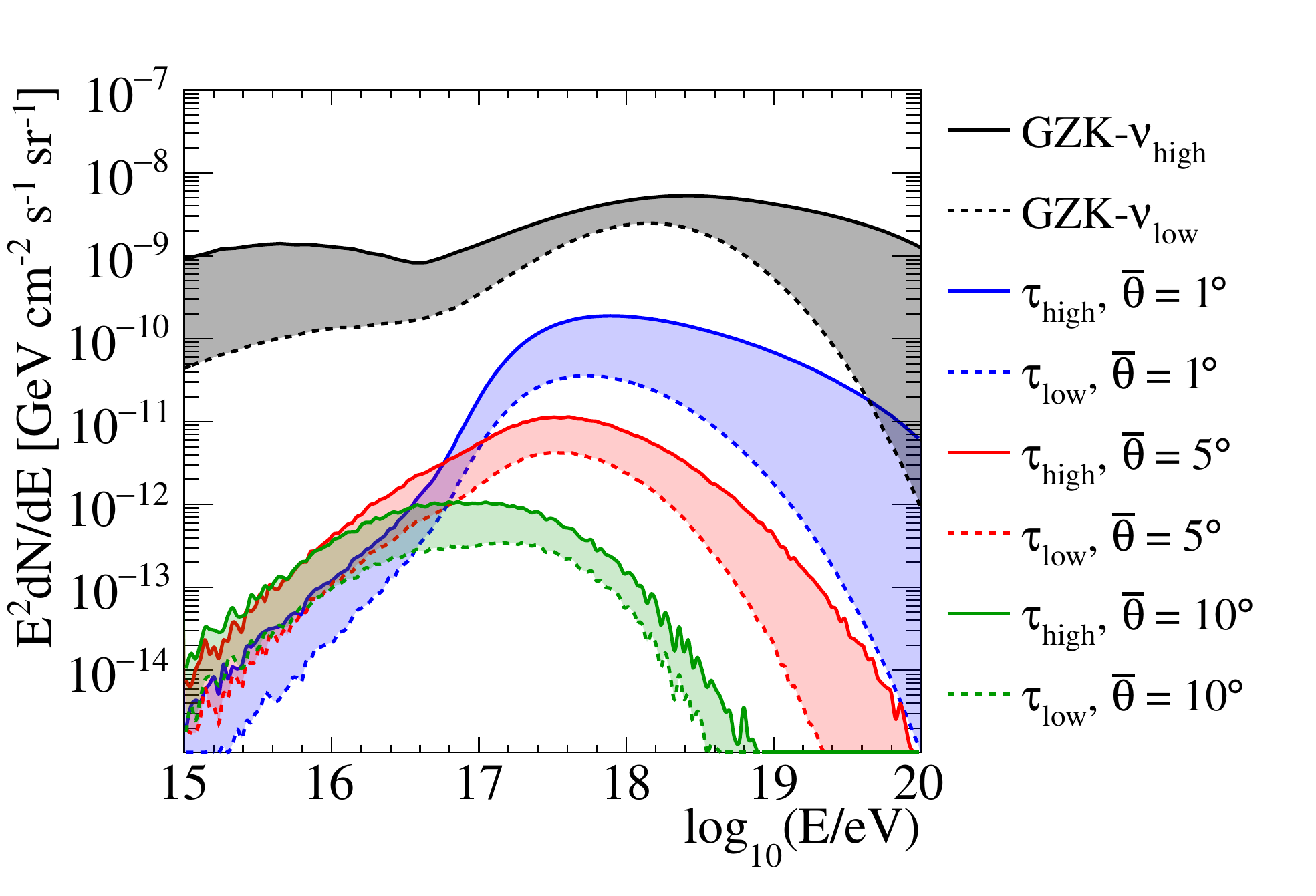} %
   \caption{The range of cosmogenic neutrino fluxes from Kotera~2010~\cite{Kotera_2010} and the resulting flux of $\tau$ leptons for emergence angles $\bar\theta=$ $1^\circ$, $5^\circ$, and $10^\circ$ (see Figure~2). The results use the middle neutrino-nucleon cross-section curve (Figure~3), ALLM energy loss rate (Figure~4) and $D=4$~km thick ice.}
   \label{fig:Kotera_tau_flux}
\end{figure}
  
\begin{figure}[!ht]
  \centering
   \includegraphics[width=1.0\linewidth]{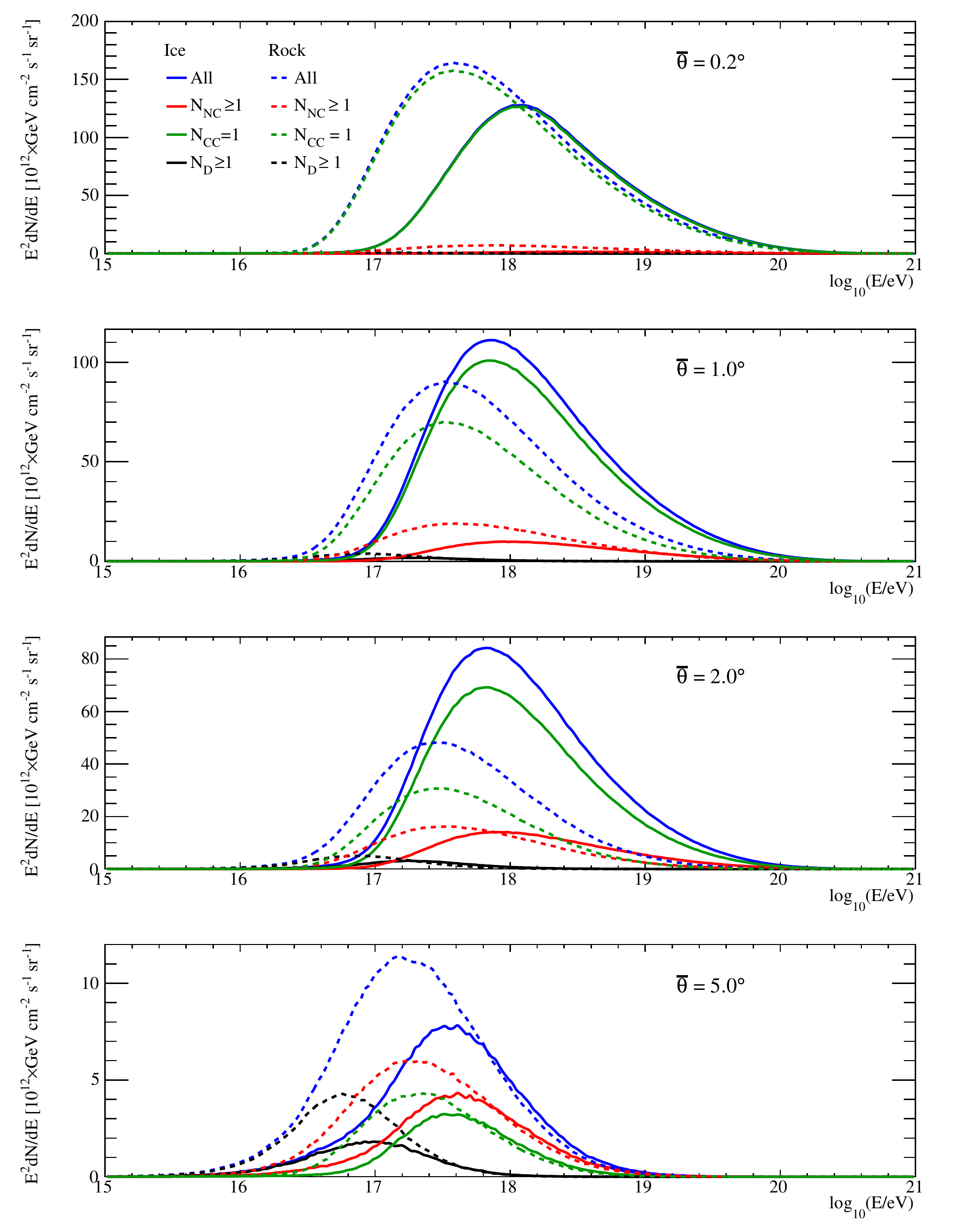} %
   \caption{The resulting flux of $\tau$ leptons for a cosmogenic neutrino flux in the middle of the flux ranges from Kotera~2010~\cite{Kotera_2010} (Grey band Figure~\ref{fig:Kotera_tau_flux}). The different line colors indicate the interaction history that led to the exiting $\tau$ leptons (see text for more details). We show the effect of a 4~km thick ice layer (solid lines) versus bare rock (dashed lines) for 4 different emergence angles as indicated on the panels. These results are obtained using the middle neutrino-nucleon cross-section curve and ALLM energy loss rate.}
   \label{fig:Channels_RockIce}
\end{figure}

\begin{figure}[!ht]
  \centering
   \includegraphics[width=1.0\linewidth]{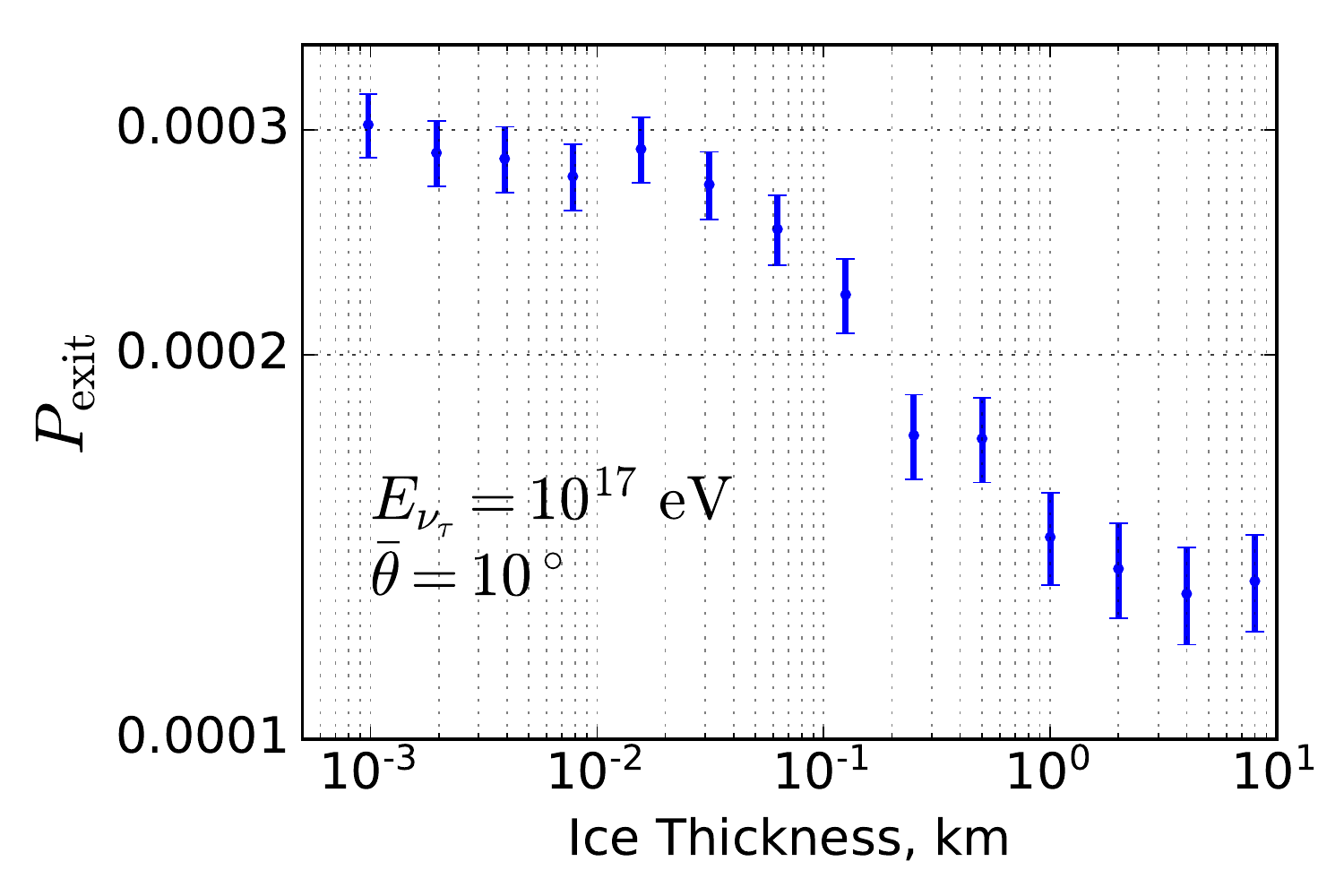}
   \caption{The $\tau$ lepton exit probability $P_\mathrm{exit}$ as a function of ice layer thickness for neutrinos with energy $E_{\nu_\tau}=10^{17}$~eV and emergence angle $\bar\theta=10^\circ$.}
   \label{fig:Ice_Thickness_Fine}
\end{figure}
\clearpage


\vspace{+11pt}
{\it Acknowledgements} Part of this work was carried out at the Jet Propulsion Laboratory, California Institute of Technology, under a contract with the National Aeronautics and Space Administration.
J. A-M and E.Z. thank Ministerio de Econom\'\i a (FPA 2015-70420-C2-1-R and FPA2017-85114-P), 
Consolider-Ingenio 2010 CPAN Programme (CSD2007-00042),  
Xunta de Galicia (GRC2013-024 and ED431C 2017/07), Feder Funds,  
$7^{\rm th}$ Framework Program (PIRSES-2009-GA-246806) 
and RENATA Red Nacional Tem\'atica de Astropart\'\i culas (FPA2015-68783-REDT). W.C. thanks grant \#2015/15735-1, S\~ao Paulo Research Foundation (FAPESP).
We thank N. Armesto and G. Parente for fruitful discussions on the neutrino cross-section and $\tau$ lepton energy-loss models.

\noindent\textsf{\copyright} 2019. All rights reserved.
\bigskip


\end{document}